\documentclass[11pt,showpacs,preprintnumbers,amsmath,amssymb,prd,nofootinbib,superscriptaddress]{revtex4-2}

\usepackage{dcolumn}
\usepackage{bm}
\usepackage{ifpdf}
\usepackage{hyperref}
\usepackage{xcolor,color,graphicx,graphics,physics}
\usepackage[spanish,english]{babel}
\usepackage[latin1]{inputenc}
\usepackage[OT1]{fontenc}
\usepackage{latexsym,amssymb,amsmath,amsfonts, slashed,cancel, simpler-wick}
\usepackage{makeidx}
\usepackage{epsfig,subfigure}
\usepackage{natbib}
\usepackage{epstopdf}
\usepackage{mathrsfs}
\usepackage{hyperref}
\hypersetup{colorlinks=true, linkcolor=blue, citecolor=blue, urlcolor=blue}
\usepackage{enumerate}
\usepackage{tikz}
\usepackage{feynmp-auto}  
\usepackage{tikz-feynman}
\tikzfeynmanset{compat=1.1.0}
\usepackage{fixmath}

\everymath{\displaystyle}
\usepackage{graphicx}

\usepackage[T1]{fontenc}
\usepackage{amsmath}
\usepackage{amssymb}
\usepackage{graphicx}
\usepackage{xcolor}

\newcommand{\bea}{\begin{eqnarray}}
\newcommand{\eea}{\end{eqnarray}}

\newcommand{\orcid}[1]{\href{https://orcid.org/#1}{\includegraphics[width=10pt]{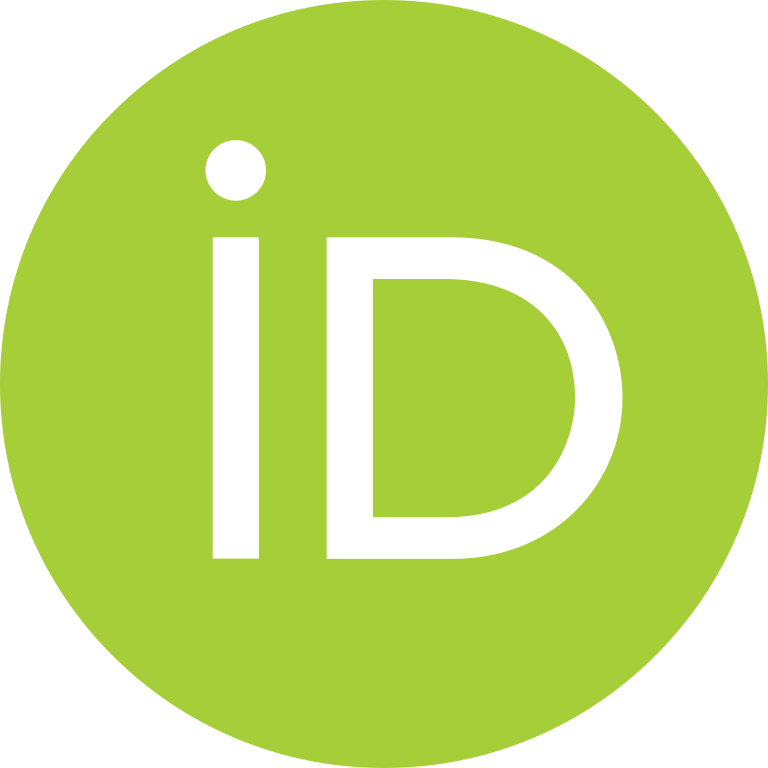}}}

\begin{document}

\title{Newtonian Potential in Weyl Gravitoelectromagnetism}

\author{L. A. S. Evangelista \orcid{0009-0002-3136-2234}}
\email{lucassouza@fisica.ufmt.br}
\affiliation{Programa de P\'{o}s-Gradua\c{c}\~{a}o em F\'{\i}sica, Instituto de F\'{\i}sica,\\ 
Universidade Federal de Mato Grosso, Cuiab\'{a}, Brasil}

\author{A. F. Santos \orcid{0000-0002-2505-5273}}
\email{alesandroferreira@fisica.ufmt.br}
\affiliation{Programa de P\'{o}s-Gradua\c{c}\~{a}o em F\'{\i}sica, Instituto de F\'{\i}sica,\\ 
Universidade Federal de Mato Grosso, Cuiab\'{a}, Brasil}

\begin{abstract}
	
    The gauge structure of the Weyl Gravitoelectromagnetic (GEM) formalism has been investigated, showing that although the theory admits additional propagating modes, only an effective spin-2 sector contributes to physical observables. Building on this result, the formalism is applied to Bhabha scattering mediated by the tensor field $A_{\mu\nu}$, and the Newtonian gravitational potential is derived in the zero-temperature limit. Finite-temperature effects are incorporated through the Thermo Field Dynamics (TFD) formalism. While the Newtonian interaction is recovered at low temperatures, it becomes progressively suppressed in the high-temperature regime, revealing a thermal screening mechanism. The physical origin of this behavior and its possible dependence on the underlying scattering process are briefly discussed.
	
\end{abstract}

\maketitle

\section{Introduction}

General Relativity (GR) \cite{einstein1916naherungsweise} is widely regarded as the fundamental theory describing the relationship between matter and the curvature of spacetime, providing the primary framework for both theoretical and phenomenological studies of the macroscopic universe. In the weak-field limit, the theory naturally recovers Newtonian gravity. Within this regime, one of the standard perturbative approaches to gravity treats the graviton as the mediator of the gravitational interaction through the Fierz-Pauli formulation of linearized GR \cite{fierz1939relativistic}. Although this theory is non-renormalizable, it possesses well-defined Feynman rules and propagators \cite{choi1993lowest}, starting from a symmetric rank-2 tensor field $h_{\mu\nu}$ propagating on Minkowski spacetime.

From the perspective of gauge field theory, linearized GR exhibits a gauge symmetry associated with infinitesimal diffeomorphism invariance, ensuring that only the physical transverse degrees of freedom propagate. This symmetry leads to the universal coupling between the spin-2 field and the energy-momentum tensor, thereby describing the interaction of gravity with external sources.

Within this context, Gravitoelectromagnetism (GEM) provides a complementary perspective. GEM is a weak-field theory that reformulates gravity in close analogy with Maxwellian electromagnetism \cite{ramos2006derivation, ramos2006differential, ramos2018weyl, ramos2020abelian}. The structural similarities between gravity and electromagnetism have been discussed since the late nineteenth century. Notable examples of theories seeking to establish such parallels include the Weyl generalization of GR \cite{weyl1929gravitation} and the Kaluza-Klein framework \cite{kaluza1921unitatsproblem, klein1926quantentheorie}. Along similar lines, GEM describes gravitational dynamics entirely in terms of quantities analogous to those of electromagnetism. The GEM formalism can be developed through three distinct approaches: (i) the direct analogy between the weak-field limit of GR and Maxwell's equations \cite{mashhoon2003gravitoelectromagnetism}, commonly referred to as linearized GEM; (ii) the decomposition of the Weyl curvature tensor into its gravitoelectric and gravitomagnetic components \cite{faqir2010lagrangian}, known as Weyl GEM; and (iii) the tidal tensor formalism \cite{costa2008gravitoelectromagnetic}. In the present work, the second approach is adopted, as its Lagrangian formulation is well established in the literature.

Within the Weyl GEM framework, the gravitational interaction is described by a massless symmetric rank-2 tensor field, $A_{\mu\nu}$. This formulation establishes a close analogy with Quantum Electrodynamics (QED), the only essential distinction being the tensorial nature of the gravitational field. In a recent work \cite{evangelista2026diffeomorphism}, the gauge structure of Weyl GEM was investigated in detail, and the Feynman propagator of the theory, which had not previously been derived explicitly, was obtained. To preserve the structural analogy with QED, a scalar-generated gauge symmetry, resembling a diffeomorphism-like transformation, is imposed on the field $A_{\mu\nu}$. This choice maintains the correspondence between the physical observables of GEM and QED while preserving a gravitational gauge interpretation. Similar restricted gauge structures have appeared in other theoretical frameworks, such as Unimodular Gravity (UG) \cite{alvarez2024primer} and Double Copy theory \cite{bern2008new}. Although these similarities motivate the present gauge choice, the precise relationship between these theories and Weyl GEM remains an open question that deserves further investigation. This scalar-generated gauge symmetry introduces additional propagation modes beyond the spin-2 sector, including spin-1 and spin-0 components. However, after an appropriate gauge fixing, the spin-1 sector becomes purely gauge, leaving an effective spin-2 theory together with an additional scalar contribution. Furthermore, our recent work demonstrated that, in physical scattering processes, the Weyl GEM formalism couples to external sources in exactly the same manner as linearized GR, reproducing the same interaction vertices with matter fields \cite{choi1993lowest}. This result highlights one of the main advantages of the Weyl GEM approach: the fundamental tensor field arises naturally from the theory itself, rather than being introduced as a perturbation of the spacetime metric. Although Weyl GEM has already been applied to several physical problems, its behavior in different backgrounds, particularly thermal environments, remains largely unexplored. This constitutes one of the main motivations of the present work, where thermal effects are incorporated through the Thermo Field Dynamics (TFD) formalism.

TFD is a real-time finite-temperature formalism developed extensively by Takahashi and Umezawa \cite{Umezawa1, Umezawa2}, becoming a well-established framework in quantum field theory. Its central idea is to represent thermal expectation values as vacuum expectation values in an enlarged Hilbert space. To accomplish this, the physical Hilbert space $\mathbb{H}$ is doubled by introducing an identical auxiliary space, the tilde space $\widetilde{\mathbb{H}}$, such that the total thermal Hilbert space is given by $\mathbb{H}_T=\mathbb{H}\otimes\widetilde{\mathbb{H}}$. Consequently, every field operator acquires a corresponding tilde partner, which is physically interpreted as describing the thermal reservoir. In a recent work \cite{cabral2026geometric}, TFD was investigated in the context of accelerated observers in de Sitter spacetime, following a close analogy with the geometric interpretation of TFD in black-hole thermodynamics originally proposed by Israel \cite{israel1976thermo}. In that study, an explicit relationship between geometric thermal effects and conventional thermodynamic temperature was established, demonstrating that the doubling of the Hilbert space is not merely a mathematical construction, but rather a natural consequence of describing the same physical system from different accelerated reference frames. These results suggest that TFD provides a particularly suitable framework for investigating thermal effects in gravitational systems.

In the present work, these recent developments in Weyl GEM are applied to Bhabha scattering in order to derive the corresponding effective gravitational potential. Subsequently, finite-temperature effects are incorporated through the TFD formalism to investigate how a thermal environment modifies the interaction between particles, providing a complementary perspective to the conventional zero-temperature analysis.

This work is organized as follows. In Section \ref{II}, the fundamental aspects and applications of the Weyl GEM formalism at zero temperature are presented. Specifically, Subsection \ref{IIA} introduces the theoretical framework of Weyl GEM, while Subsection \ref{IIB} provides the explicit derivation of the Newtonian effective potential. In Section \ref{III}, the TFD formalism is reviewed. Subsection \ref{IIIA} presents its theoretical foundations, whereas Subsection \ref{IIIB} derives the finite-temperature Feynman propagator of the $A_{\mu\nu}$ field. In Section \ref{IV}, the tools developed in the previous sections are employed to derive the finite-temperature effective potential, followed by a detailed discussion of the results. Finally, the conclusions are presented in Section \ref{V}.

\section{Gravitoelectromagnetism at Zero Temperature}\label{II}

This section is devoted to the calculation of the gravitational potential within the Weyl Gravi\-toelectromagnetic (GEM) formalism. To this end, Bhabha scattering, $e^-+e^+\rightarrow e^-+e^+$, is considered. Before presenting the calculation, the main aspects of the Weyl GEM theory are briefly reviewed.

\subsection{The Weyl GEM formalism}\label{IIA}

Weyl GEM is one of the three main approaches to Gravitoelectromagnetism. Its formulation is based on the decomposition of the Weyl curvature tensor into its gravitoelectric and gravitomagnetic components. Alternatively, Weyl GEM can be formulated using group-theoretical methods, through which its fundamental concepts provide a well-defined effective spin-2 field theory \cite{faqir2010lagrangian}. Within this framework, a consistent Lagrangian formulation exists, allowing the theory to be quantized. The Weyl GEM Lagrangian density is given by
\begin{eqnarray}
	\mathcal{L}_{\text{GEM}} = -\frac{1}{16\pi} F_{\mu\nu\alpha} F^{\mu\nu\alpha} + G \mathcal{J}^{\nu\alpha} A_{\nu\alpha},
\end{eqnarray}
where $F^{\mu\nu\alpha} = \partial^\mu A^{\nu\alpha} - \partial^\nu A^{\mu\alpha}$ is the GEM field-strength tensor, $A_{\mu\nu}$ is the symmetric tensor potential that constitutes the fundamental field of the theory, $G$ is the gravitational constant, and $\mathcal{J}^{\nu\alpha}$ is a second-rank source tensor constructed from the mass density and mass-current density, representing the sources of the gravitational field. The corresponding field equations are
\begin{align}
	\partial_{\mu} F^{\mu\nu\alpha} &= -\frac{4\pi G}{c}\mathcal{J}^{\nu\alpha}, \label{tensorF}\\
	\partial_{\mu} G^{\mu\nu\alpha} &= 0, \label{tensorG}
\end{align}
where $G^{\mu\nu\alpha}$ denotes the dual GEM field tensor, in direct analogy with electromagnetism. In terms of the potential tensor $A_{\mu\nu}$, the field equation in the presence of sources becomes
\begin{eqnarray}
	\Box A^{\nu\alpha} - \partial^\nu(\partial_\mu A^{\mu\alpha}) = 4\pi G \mathcal{J}^{\nu\alpha}.
\end{eqnarray}
Owing to the gauge invariance of the GEM potential, analogous to the electromagnetic case, one may impose the Lorenz-type gauge condition $\partial_\mu A^{\mu\alpha} = 0$, reducing the field equation to
\begin{eqnarray}
	\Box A^{\nu\alpha} = 4\pi G \mathcal{J}^{\nu\alpha}.
\end{eqnarray}
This equation closely resembles the linearized Einstein field equations written in terms of the trace-reversed metric perturbation,
\begin{eqnarray}
	\Box \bar{h}_{\mu\nu} = 16\pi G T_{\mu\nu},
\end{eqnarray}
where $\bar{h}_{\mu\nu} = h_{\mu\nu} - \frac{1}{2}\eta_{\mu\nu} h$ \cite{misner1973k}. The structural similarity between these equations highlights the correspondence between GEM and linearized GR. In particular, the tensor $A^{\nu\alpha}$ in the Weyl GEM formalism encodes the essential dynamical content of the theory, playing a role analogous to that of the metric perturbation in the weak-field limit. For a more detailed discussion of the similarities and differences between GEM and GR, see Refs.~\cite{bakopoulos2016gravitoelectromagnetism, mashhoon2003gravitoelectromagnetism, alesandrogravitacional}.

The GEM field transforms under a gauge symmetry according to
\begin{equation}
	A_{\mu\nu}
	\longrightarrow
	A_{\mu\nu}
	+
	\partial_\mu \theta_\nu.
\end{equation}
In general, the symmetry of the tensor field would require a symmetrized transformation. Instead of introducing this symmetrization explicitly, the following condition is imposed:
\begin{align}
	\partial_\mu \theta_\nu = \partial_\nu \theta_\mu,
\end{align}
which implies that $\theta_\nu$ can be written as a pure gradient,
\begin{equation}
	\theta_\nu = \partial_\nu \lambda,
\end{equation}
where $\lambda$ is a scalar function. This condition ensures the symmetry of the tensor field while restricting the number of independent gauge degrees of freedom. It also establishes a formal analogy with electrodynamics, in which the gauge symmetry is controlled by a scalar parameter. As a consequence, the resulting theory does not correspond to a fully general spin-2 gauge theory, but rather to a tensorial theory with an effectively induced scalar gauge symmetry that reproduces gravitational interactions in physical processes. This scalar-generated gauge symmetry plays a central role in ensuring the consistency of the model and distinguishes it from full diffeomorphism invariance. As a consequence of this gauge structure, the theory admits both spin-2 and spin-0 propagating modes, leading to the following Feynman propagator for the $A_{\mu\nu}$ field:
\begin{align}
	D_{\mu\nu,\rho\sigma}(q)=\frac{1}{2q^2}\left(\eta_{\mu\rho}\eta_{\nu\sigma}+\eta_{\mu\sigma}\eta_{\nu\rho}\right).\label{propagatorgem}
\end{align}
It is worth noting that this propagator has the same structure as the graviton propagator in linearized GR, except for the absence of the third metric combination. In previous applications of the GEM theory \cite{alesandrogravitacional, evangelista2026gravitational, santos2017gravitational, Casimir}, the linearized GR graviton propagator was adopted. However, in a more recent work, the propagator of the Weyl GEM theory was explicitly derived, yielding Eq.~\eqref{propagatorgem}. Nevertheless, both propagators produce identical physical scattering amplitudes, so the results reported in the previous literature remain unchanged despite the different propagator structure.

For a detailed discussion of the choice of gauge symmetry for the $A_{\mu\nu}$ field, the motivation for this particular construction, and the explicit derivation of the Feynman propagator, including the emergence of the additional spin modes, we refer the reader to Ref.~\cite{evangelista2026diffeomorphism}.
In the next subsection, the formalism presented here will be employed to analyze the effective gravitational potential associated with Bhabha scattering, where electrons exchange virtual $A_{\mu\nu}$ quanta at zero temperature.

\subsection{The effective potential zero temperature }\label{IIB}

\begin{figure}[!h]
	\centering
	\includegraphics[width=0.7\linewidth]{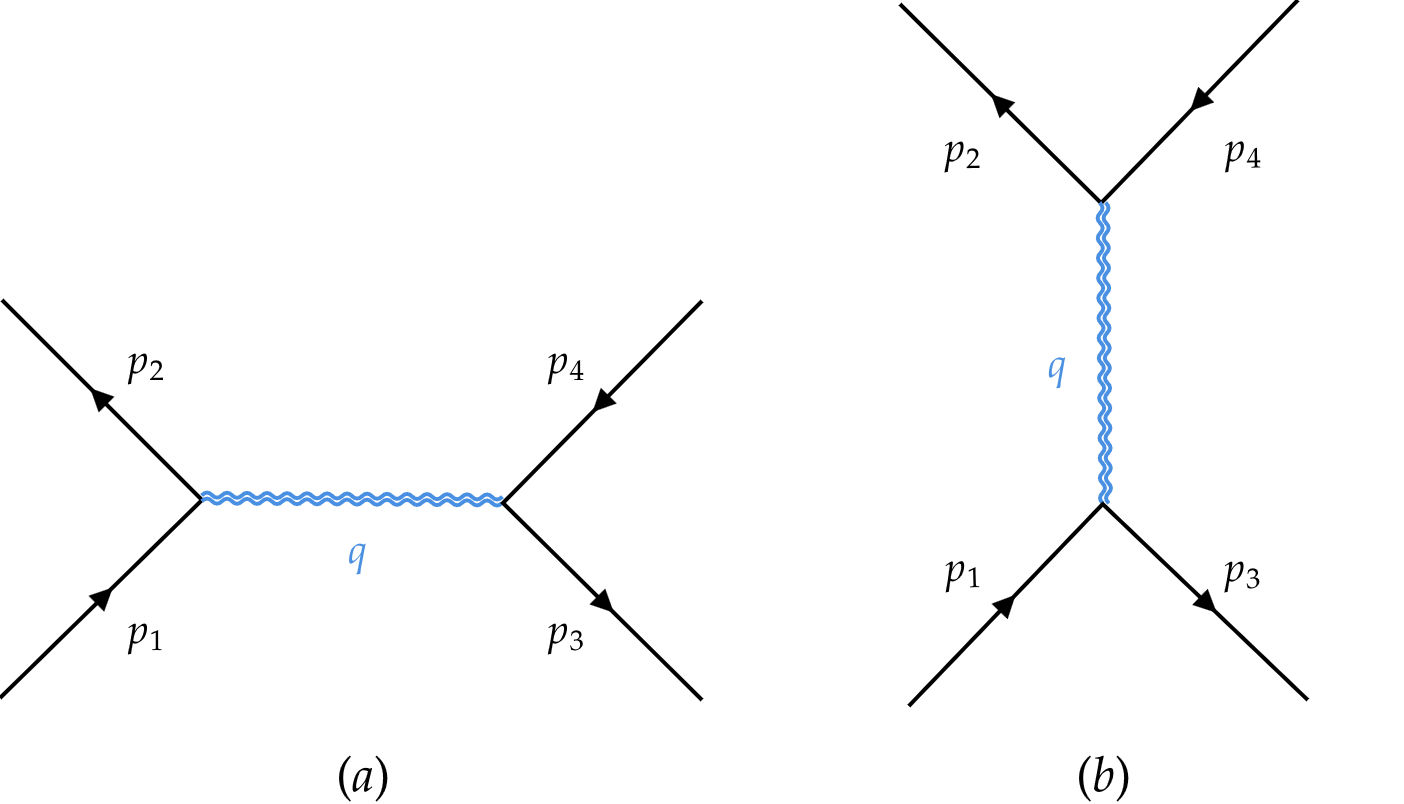}
	\caption{Feynman diagrams contributing to Bhabha scattering. The diagram (a) corresponds to the $s$-channel, while the diagram (b) corresponds to the $t$-channel. Time flows from left to right.}
	\label{diagram}
\end{figure}

This subsection is dedicated to the calculation of the Bhabha scattering process, $e^-+e^+\to e^-+e^+$, represented by the Feynman diagrams in Fig. \ref{diagram}.

The interaction Lagrangian describing the coupling between the $A_{\mu\nu}$ field and the fermionic field is given by
\begin{align}
	\mathcal{L}_{A\psi\psi}=\frac{i\kappa}{4}A_{\mu\nu}\overline{\psi}\left(\gamma^\mu \overleftrightarrow{\partial^\nu}+\gamma^\nu \overleftrightarrow{\partial^\mu}\right)\psi,
\end{align}
where $\psi$ denotes the Dirac field, with $\overline{\psi}=\psi^\dagger \gamma^0$, and $\kappa=\sqrt{4\pi G}$ is the GEM coupling constant. This interaction leads to the following vertex factor:
\begin{align}
	V^{\mu\nu}_{A\psi\psi}=\frac{i\kappa}{4}\left[\gamma^\mu\left(p+p'\right)^\nu+\gamma^\nu\left(p+p'\right)^\mu\right],
\end{align}
where $p$ and $p'$ are the incoming and outgoing fermion four-momenta, respectively. The physical contribution to the scattering amplitude is given by the two tree-level Feynman diagrams, yielding
\begin{align}
	\mathcal{M}=\overline{u}(p_2)V^{\mu\nu}_{A\psi\psi}v(p_1)D_{\mu\nu,\rho\sigma}(q)\overline{v}(p_3)V^{\rho\sigma}_{A\psi\psi}u(p_4)-\overline{u}(p_3)V^{\mu\nu}_{A\psi\psi}u(p_1)D_{\mu\nu,\rho\sigma}(q)\overline{v}(p_2)V^{\rho\sigma}_{A\psi\psi}v(p_4),
\end{align}
where the first and second terms correspond to the $s$-channel and $t$-channel diagrams, respectively.

Since the objective is to derive the effective interaction potential from the scattering amplitude, only the $t$-channel contribution is considered, as it is the only one that contributes to the non-relativistic potential. Thus,
\begin{align}
	\mathcal{M}=-\overline{u}(p_3)V^{\mu\nu}_{A\psi\psi}u(p_1)D_{\mu\nu,\rho\sigma}(q)\overline{v}(p_2)V^{\rho\sigma}_{A\psi\psi}v(p_4).
\end{align}
In the non-relativistic limit and in the center-of-mass frame, we have \cite{peskin2018introduction}
\begin{align}
	p_1=(m,\Vec{p}_1), \quad p_2=(m,\Vec{p}_2), \quad p_3=(m,\Vec{p}_3), \quad p_4=(m,\Vec{p}_4),
\end{align}
together with
\begin{align}
	(p_1-p_3)^2\approx -|\Vec{p}_1-\Vec{p}_3|^2,
\end{align}
where $p_i$, with $i=1,...,4$, denote the four-momenta of the incoming and outgoing fermions.
In this regime, the transferred four-momentum $q$ satisfies $q\approx(0,-\Vec{q})$. Furthermore,
\begin{align}
	\overline{u}(p_3)\gamma^0u(p_1)&=u^\dagger(p_3)u(p_1)\approx2m\,\delta_{s_3,s_1},\\
	\overline{v}(p_2)\gamma^0v(p_4)&=v^\dagger(p_2)v(p_4)\approx2m\,\delta_{s_2,s_4}
\end{align}
with $s_i$ denoting the spin indices.
Since $|\Vec{p}|\ll m$, the temporal components dominate the four-momenta ($p^0\approx m$). Consequently, the leading contribution to the interaction vertex is the temporal component, i.e.,
\begin{align}
	V^{00}=i\kappa m\gamma^0,
\end{align}
which naturally selects the $\gamma^0$ matrix in the spinor bilinears.

With these approximations, the scattering amplitude reduces to
\begin{align}
	\mathcal{M}=\frac{4\kappa^2m^4}{\Vec{q}^{\,2}}.
\end{align}
The effective potential is then obtained within the Born approximation as
\begin{align}
	V(\Vec{r})=-\int\frac{d^3\Vec{q}}{(2\pi)^3}\left(\frac{\mathcal{M}}{4m^2}\right)e^{i\Vec{q}\cdot\Vec{r}},
\end{align}
which yields
\begin{align}
	V(\Vec{r})=-\frac{\kappa^2m^2}{4\pi r}=-\frac{Gm^2}{r}.
\end{align}
It is important to note that the choice $\kappa=\sqrt{4\pi G}$ is required to recover the Newtonian effective potential. This normalization differs from that commonly adopted in linearized GR \cite{choi1993lowest}. The difference originates from the propagation of an additional scalar mode in the Weyl GEM formalism, which modifies the normalization of the interaction while leaving the physical scattering amplitude unchanged. Consequently, the choice of $\kappa=\sqrt{4\pi G}$ provides the correct Newtonian limit within the present framework.

Therefore, the Newtonian gravitational potential is explicitly recovered within the Weyl GEM formalism at zero temperature. This result highlights the effectiveness of the formalism in describing gravitational interactions through standard quantum field-theoretical techniques. In particular, the calculation of Feynman rules, propagators, and scattering amplitudes is considerably more straightforward than in linearized General Relativity, where the geometric structure of the metric perturbations leads to a significantly more involved perturbative framework.

In the next section, the main aspects of the Thermo Field Dynamics formalism, which is employed to incorporate thermal effects into the theory, will be presented. This will be followed by the explicit derivation of the finite-temperature propagator for the $A_{\mu\nu}$ field.

\section{Thermo Field Dynamics and the GEM Propagator}\label{III}

In this section, the Thermo Field Dynamics (TFD) formalism at finite temperature is introduced. The discussion is organized into two parts. The first reviews the essential features of TFD, while the second applies the formalism to derive the finite-temperature Feynman propagator for the GEM field. 

\subsection{Thermo Field Dynamics framework}\label{IIIA}

TFD is a real-time formalism based on the doubling of the Hilbert space $\mathbb{H}$ into an identical copy, $\widetilde{\mathbb{H}}$, such that the thermal Hilbert space is defined as $\mathbb{H}_T=\mathbb{H}\otimes\widetilde{\mathbb{H}}$. At first sight, one might expect that this doubling introduces unphysical degrees of freedom into the theory, leading to duplicated contributions in physical observables. Although the doubled space indeed contains auxiliary degrees of freedom, the TFD formalism provides a systematic prescription for isolating only the physical sector, ensuring that all observable quantities remain unaffected by the unphysical contributions \cite{Book}.

To construct an operator $\mathcal{O}$ in the doubled Hilbert space, one introduces the tilde-conjugation rules,
\begin{align}
	\widetilde{\mathcal{O}_i\mathcal{O}_j}=\widetilde{\mathcal{O}}_i\widetilde{\mathcal{O}}_j;\quad\quad
	\widetilde{\widetilde{\mathcal{O}}}=\pm\mathcal{O};\quad\quad
	\widetilde{\left(c\,\mathcal{O}_i+\mathcal{O}_j\right)}=c^*\widetilde{\mathcal{O}}_i+\widetilde{\mathcal{O}}_j;\quad\quad
	\widetilde{\mathcal{O}^\dagger}=\left(\widetilde{\mathcal{O}}\right)^\dagger,
\end{align}
where the sign $\pm$ depends on the statistics of the field, being positive for bosons and negative for fermions, while $c$ is a complex constant. In addition, the doubled operators satisfy the algebra
\begin{align}
	\left[\mathcal{O}_i,\mathcal{O}_j\right]=i\varepsilon^k_{ij}\mathcal{O}_k;\quad\quad
	\left[\widetilde{\mathcal{O}}_i,\widetilde{\mathcal{O}}_j\right]=-i\varepsilon^k_{ij}\widetilde{\mathcal{O}}_k;\quad\quad
	\left[\mathcal{O}_i,\widetilde{\mathcal{O}}_j\right]=0,
\end{align}
where $\varepsilon^k_{ij}$ denotes the structure constants of the algebra.

The thermal expectation value of the observable $\mathcal{O}$ is defined as
\begin{align}
	\langle\mathcal{O}\rangle_\beta=\bra{0(\beta)}\mathcal{O}\ket{0(\beta)},
\end{align}
where the thermal vacuum is given by
\begin{align}
	\ket{0(\beta)}=\frac{1}{\sqrt{Z(\beta)}}\sum_n e^{-\beta E_n/2}\ket{n,\tilde{n}},
\end{align}
with $\beta=1/(k_B T)$, where $k_B$ is the Boltzmann constant, and $Z(\beta)$ denotes the partition function. The doubled representation of an operator is then defined as
\begin{align}
	\mathcal{O}^a=
	\begin{pmatrix}
		\mathcal{O}\\
		\widetilde{\mathcal{O}}^\dagger
	\end{pmatrix},
\end{align}
where $a=1,2$ labels the physical (non-tilde) and tilde sectors, respectively.

The operator $\mathcal{O}$ can be expressed in the thermal basis through the Bogoliubov transformation
\begin{align}
	\mathcal{O}^a=\mathbb{M}^{ab}(\beta)\mathcal{O}_b(\beta),
\end{align}
where $\mathbb{M}^{ab}(\beta)$ is the Bogoliubov matrix. For fermions,
\begin{align}
	\mathbb{M}_F^{ab}(\beta)=
	\begin{pmatrix}
		U(\beta)&V(\beta)\\
		-V(\beta)&U(\beta)
	\end{pmatrix},
	\label{matrixfermions}
\end{align}
with $U^2(\beta)=1-f(\beta)$ and $V^2(\beta)=f(\beta)$, where $f(\beta)$ is the Fermi-Dirac distribution. For bosons,
\begin{align}
	\mathbb{M}_B^{ab}(\beta)=
	\begin{pmatrix}
		U'(\beta)&V'(\beta)\\
		V'(\beta)&U'(\beta)
	\end{pmatrix},
	\label{matrixbogubos}
\end{align}
where $U'^2(\beta)=1+n(\beta)$ and $V'^2(\beta)=n(\beta)$, with $n(\beta)$ denoting the Bose-Einstein distribution. These distributions are given by
\begin{align}
	f(\beta)=\frac{1}{e^{\beta p_0}+1},\qquad
	n(\beta)=\frac{1}{e^{\beta k_0}-1}.
\end{align}

The same formalism applies to the creation and annihilation operators. Through the Bogoliubov transformations, these operators are rotated in the doubled Hilbert space according to
\begin{align}
	b_{s,p} &= U(\beta)b_{s,p}(\beta)+V(\beta)\widetilde{b}_{s,p}^\dagger(\beta),\quad\quad
	\widetilde{b}_{s,p}=U(\beta)\widetilde{b}_{s,p}(\beta)-V(\beta)b_{s,p}^\dagger(\beta),\nonumber\\
	b_{s,p}^\dagger &= U(\beta)b_{s,p}^\dagger(\beta)+V(\beta)\widetilde{b}_{s,p}(\beta),\quad\quad
	\widetilde{b}_{s,p}^\dagger=U(\beta)\widetilde{b}_{s,p}^\dagger(\beta)-V(\beta)b_{s,p}(\beta),
	\label{boguferm}
\end{align}
for fermions, and
\begin{align}
	a_{\lambda,k} &= U'(\beta)a_{\lambda,k}(\beta)+V'(\beta)\widetilde{a}_{\lambda,k}^\dagger(\beta),\quad\quad
	\widetilde{a}_{\lambda,k}=U'(\beta)\widetilde{a}_{\lambda,k}(\beta)+V'(\beta)a_{\lambda,k}^\dagger(\beta),\nonumber\\
	a_{\lambda,k}^\dagger &= U'(\beta)a_{\lambda,k}^\dagger(\beta)+V'(\beta)\widetilde{a}_{\lambda,k}(\beta),\quad\quad
	\widetilde{a}_{\lambda,k}^\dagger=U'(\beta)\widetilde{a}_{\lambda,k}^\dagger(\beta)+V'(\beta)a_{\lambda,k}(\beta),
	\label{bogubos}
\end{align}
for bosons, where $s$ and $\lambda$ denote the spin and polarization indices, respectively.

The fermionic operators satisfy the anticommutation relations
\begin{align}
	\left\{b_{s,p}(\beta),b^\dagger_{s',p'}(\beta)\right\}
	=
	\left\{\widetilde{b}_{s,p}(\beta),\widetilde{b}^\dagger_{s',p'}(\beta)\right\}
	=
	(2\pi)^3\delta^3(p-p')\delta_{s,s'},
\end{align}
whereas the bosonic operators satisfy
\begin{align}
	\left[a_{\lambda,k}(\beta),a^\dagger_{\lambda',k'}(\beta)\right]
	=
	\left[\widetilde{a}_{\lambda,k}(\beta),\widetilde{a}^\dagger_{\lambda',k'}(\beta)\right]
	=
	(2\pi)^3\delta^3(k-k')\delta_{\lambda,\lambda'}.
	\label{commutationrelation}
\end{align}
All remaining (anti)commutators vanish.

The discussion presented above provides the essential ingredients required to describe particle scattering within the TFD formalism. Since Bhabha scattering mediated by the Gravitoelectromagnetic field is considered, the fermionic Bogoliubov transformations given in Eq.~\eqref{matrixfermions} are employed. However, the construction of the finite-temperature GEM propagator requires additional ingredients.

With this motivation, the next subsection introduces the formalism needed to derive the thermal corrections to the propagator of the $A_{\mu\nu}$ field.

\subsection{Finite-temperature GEM propagator}\label{IIIB}

In order to calculate the Feynman propagator for the GEM theory at finite temperature, the matrix formulation is adopted rather than the usual vacuum expectation value approach. This choice is motivated by the fact that the thermal contribution to the GEM field propagator has already been derived in the literature \cite{graviton}. Since our zero-temperature propagator now contains an additional scalar contribution, we rederive it here for completeness using a different formalism that has not been discussed in previous works. Furthermore, it is shown that the sum over polarization tensors reproduces the propagator structure given in Eq.~\eqref{propagatorgem}.

With this in mind, the $A_{\mu\nu}$ field can be expanded as the plane-wave solution
\begin{align}
	A_{\mu\nu}(x)=\int \frac{d^3k}{(2\pi)^3}\frac{1}{\sqrt{2 k_0}}\sum_{\lambda}\epsilon_{\mu\nu}^{(\lambda)}\left(a_{\lambda,k}e^{-ikx}+a^\dagger_{\lambda,k}e^{ikx}\right),
\end{align}
where $\epsilon_{\mu\nu}^{(\lambda)}$ is the GEM polarization tensor. By definition, the zero-temperature Feynman propagator in the matrix representation \cite{Book} is given by
\begin{align}
	D^{ab}_{\mu\nu,\rho\sigma}(q)=M_{\mu\nu,\rho\sigma}^{(\lambda)}\Delta_{0}^{ab}(q),
\end{align}
with
\begin{align}
	M_{\mu\nu,\rho\sigma}^{(\lambda)}=\sum_{\lambda}\epsilon_{\mu\nu}^{(\lambda)}\epsilon_{\rho\sigma}^{*(\lambda)},\quad\quad
	\Delta_{0}^{ab}(q)=\begin{pmatrix}
		\frac{1}{q^2+i\eta}&0\\
		0&-\frac{1}{q^2-i\eta}
	\end{pmatrix},
\end{align}
where the commutation relation \eqref{commutationrelation} has been used. To introduce thermal effects, the Bogoliubov transformation is applied
\begin{align}
	\Delta^{ab}_\beta(q)=\mathbb{M}_B^{ab}(\beta)\Delta_0^{ab}(q)\mathbb{M}_B^{ab}(\beta).\label{Deltabeta}
\end{align}
In hyperbolic form, the bosonic Bogoliubov transformation in Eq.~\eqref{matrixbogubos} can be written as
\begin{align}
	\mathbb{M}_B^{ab}=\begin{pmatrix}
		\cosh\theta_{\beta}&\sinh\theta_\beta\\
		\sinh\theta_\beta&\cosh\theta_\beta
	\end{pmatrix},
\end{align}
which can be substituted into Eq.~\eqref{Deltabeta} to obtain
\begin{align}
	\Delta^{ab}_\beta(q)&=\begin{pmatrix}
		\frac{\cosh^2\theta_\beta}{q^2+i\eta}-\frac{\sinh^2\theta_\beta}{q^2-i\eta}&
		\frac{\cosh\theta_\beta\sinh\theta_\beta}{q^2+i\eta}-\frac{\cosh\theta_\beta\sinh\theta_\beta}{q^2-i\eta}\\
		\frac{\cosh\theta_\beta\sinh\theta_\beta}{q^2+i\eta}-\frac{\cosh\theta_\beta\sinh\theta_\beta}{q^2-i\eta}&
		\frac{\sinh^2\theta_\beta}{q^2+i\eta}-\frac{\cosh^2\theta_\beta}{q^2-i\eta}
	\end{pmatrix}.
\end{align}
Here, the cutting rules can be used, namely
\begin{align}
	\frac{1}{q^2\pm i\eta}=\frac{1}{q^2}\mp i\pi \delta(q^2),
\end{align}
which yield the thermal propagator
\begin{align}
	\Delta^{ab}_\beta(q)=\begin{pmatrix}
		\frac{1}{q^2+i\eta}-2\pi in(\beta)\delta(q^2)&
		-2\pi i\sqrt{n(\beta)(1+n(\beta))}\delta(q^2)\\
		-2\pi i\sqrt{n(\beta)(1+n(\beta))}\delta(q^2)&
		-\frac{1}{q^2-i\eta}-2\pi in(\beta)\delta(q^2)
	\end{pmatrix}.\label{TFDMatrix}
\end{align}

Turning to the tensor $M_{\mu\nu,\rho\sigma}^{(\lambda)}$, it is noted that, for a symmetric helicity-$2\lambda$ polarization tensor ($\lambda=\pm1$), it can be expressed as the product of two polarization vectors \cite{choi1993lowest},
\begin{align}
	\epsilon_{\mu\nu}^{(2\lambda)}=\epsilon_\mu^{(\lambda)}\epsilon_\nu^{(\lambda)}.
\end{align}
Furthermore, since the polarization tensor is symmetric, it can be written as
\begin{align}
	\epsilon_\mu^{(\lambda)}\epsilon_\nu^{(\lambda)}
	=\frac{1}{2}\left(\epsilon_\mu^{(\lambda)}\epsilon_\nu^{(\lambda)}
	+\epsilon_\nu^{(\lambda)}\epsilon_\mu^{(\lambda)}\right).
\end{align}
Substituting this expression into $M_{\mu\nu,\rho\sigma}^{(\lambda)}$, we obtain
\begin{align}
	M_{\mu\nu,\rho\sigma}^{(\lambda)}
	=\frac{1}{4}\sum_{\lambda}
	\left[
	\left(\epsilon_\mu^{(\lambda)}\epsilon_\nu^{(\lambda)}
	+\epsilon_\nu^{(\lambda)}\epsilon_\mu^{(\lambda)}\right)
	\left(\epsilon_\rho^{*(\lambda)}\epsilon_\sigma^{*(\lambda)}
	+\epsilon_\sigma^{*(\lambda)}\epsilon_\rho^{*(\lambda)}\right)
	\right].
\end{align}
Expanding the products and using the completeness relation for the polarization vectors,
\begin{align}
	\sum_\lambda\epsilon_\mu^{(\lambda)}\epsilon_\rho^{*(\lambda)}
	=\eta_{\mu\rho},
\end{align}
it is found that
\begin{align}
	M_{\mu\nu,\rho\sigma}^{(\lambda)}
	=\frac{1}{2}\left(\eta_{\mu\rho}\eta_{\nu\sigma}
	+\eta_{\mu\sigma}\eta_{\nu\rho}\right),\label{Mprop}
\end{align}
which is precisely the tensor structure appearing in the GEM propagator of Eq.~\eqref{propagatorgem}. By combining Eqs.~\eqref{Mprop} and \eqref{TFDMatrix}, the finite-temperature propagator of the $A_{\mu\nu}$ field is obtained,
\begin{align}
	D^{ab}_{\mu\nu,\rho\sigma}(q;\beta)=
	\frac{1}{2}
	\left(\eta_{\mu\rho}\eta_{\nu\sigma}
	+\eta_{\mu\sigma}\eta_{\nu\rho}\right)
	\begin{pmatrix}
		\frac{1}{q^2+i\eta}-2\pi in(\beta)\delta(q^2)&
		-2\pi i\sqrt{n(\beta)(1+n(\beta))}\delta(q^2)\\
		-2\pi i\sqrt{n(\beta)(1+n(\beta))}\delta(q^2)&
		-\frac{1}{q^2-i\eta}-2\pi in(\beta)\delta(q^2)
	\end{pmatrix}.
	\label{temperaturepropagator}
\end{align}
It should be noted that the component $a=b=1$ corresponds to the physical (non-tilde) Feynman propagator. Furthermore, in the limit $n(\beta)\to0$, corresponding to $\beta\to\infty$ and therefore $T\to0$, the usual GEM propagator given in Eq.~\eqref{propagatorgem} is recovered. This propagator has the same physical structure as the one derived in Ref.~\cite{graviton}; however, the present derivation is considerably more direct.

In the next section, the formalism developed here will be applied to derive the complete finite-temperature effective potential.

\section{The Thermal Effective Potential}\label{IV}

Here, Bhabha scattering mediated by the GEM field at finite temperature is investigated. Within the TFD formalism, the interaction Lagrangian is defined as
\begin{align}
	\hat{\mathcal{L}}_{A\psi\psi}=\mathcal{L}_{A\psi\psi}-\tilde{\mathcal{L}}_{A\psi\psi},
\end{align}
where the tilde interaction Lagrangian is given by
\begin{align}
	\Tilde{\mathcal{L}}_{A\psi\psi}=\frac{i\kappa}{4}\Tilde{A}_{\mu\nu}\tilde{\overline{\psi}}\left(\gamma^{*\mu} \overleftrightarrow{\partial^\nu}+\gamma^{*\nu} \overleftrightarrow{\partial^\mu}\right)\Tilde{\psi},
\end{align}
in which the tilde operation acts on the particle fields by transforming the Dirac spinors, as well as the creation and annihilation operators contained in them, into their tilde counterparts, while taking the complex conjugate of the remaining quantities. Furthermore, the tilde matrices satisfy the same Clifford algebra as the ordinary Dirac matrices, namely $\{\gamma^\mu,\gamma^\nu\}=\{\gamma^{*\mu},\gamma^{*\nu}\}=2g^{\mu\nu}$.

Substituting the field expansions and applying the Bogoliubov transformations for the fermionic operators, given in Eq.~\eqref{boguferm}, while retaining only the $t$-channel contribution, yields the following transition amplitude for the Bhabha scattering process:
\begin{align}
	\mathcal{M}_\beta=&-U^4_\beta\left[\overline{u}(p_3)V^{\mu\nu}_{A\psi\psi}u(p_1)\right]D^{(11)}_{\mu\nu,\rho\sigma}(q;\beta)\left[\overline{v}(p_2)V^{\rho\sigma}_{A\psi\psi}v(p_4)\right]\nonumber\\
	&-U^2_\beta V^2_\beta\left[\overline{u}(p_3)V^{\mu\nu}_{A\psi\psi}u(p_1)\right]D^{(12)}_{\mu\nu,\rho\sigma}(q;\beta)\left[\tilde{\overline{v}}(p_2)\tilde{V}^{\rho\sigma}_{A\psi\psi}\tilde{v}(p_4)\right]\nonumber\\
	&-U^2_\beta V^2_\beta\left[\tilde{\overline{u}}(p_3)\tilde{V}^{\mu\nu}_{A\psi\psi}\tilde{u}(p_1)\right]D^{(21)}_{\mu\nu,\rho\sigma}(q;\beta)\left[\overline{v}(p_2)V^{\rho\sigma}_{A\psi\psi}v(p_4)\right]\nonumber\\
	&-V^4_\beta\left[\tilde{\overline{u}}(p_3)\tilde{V}^{\mu\nu}_{A\psi\psi}\tilde{u}(p_1)\right]D^{(22)}_{\mu\nu,\rho\sigma}(q;\beta)\left[\tilde{\overline{v}}(p_2)\tilde{V}^{\rho\sigma}_{A\psi\psi}\tilde{v}(p_4)\right],
\end{align}
where each term corresponds to one of the sectors of the doubled Hilbert space $\mathbb{H}_T$.

It is important to note that the non-relativistic limit of the Dirac spinors remains unchanged within the TFD formalism \cite{Book}. Therefore,
\begin{align}
	\tilde{\overline{u}}(p_3)\gamma^0 \tilde{u}(p_1)&=\overline{u}(p_3)\gamma^0 u(p_1)\approx 2m \delta_{s_3,s_1},\\
	\tilde{\overline{v}}(p_2)\gamma^0 \tilde{v}(p_4)&=\overline{v}(p_2)\gamma^0 v(p_4)\approx 2m \delta_{s_2,s_4},
\end{align}
which allows the transition amplitude to be written in the more compact form
\begin{align}
	\mathcal{M}_\beta=&-\left[\overline{u}(p_3)V^{\mu\nu}_{A\psi\psi}u(p_1)\right]\left[\overline{v}(p_2)V^{\rho\sigma}_{A\psi\psi}v(p_4)\right]\nonumber\\
	&\times\left[(U^4_\beta-V^4_\beta)D^{(11)}_{\mu\nu,\rho\sigma}(q)+(U^4_\beta+V^4_\beta)\mathbb{D}^{(11)}_{\mu\nu,\rho\sigma}(q;\beta)+2U^2_\beta V^2_\beta D^{(12)}_{\mu\nu,\rho\sigma}(q;\beta)\right],\label{Mfinal}
\end{align}
where $D^{(11)}_{\mu\nu,\rho\sigma}(q)$ denotes the zero-temperature part of the propagator. The relation
$D^{(22)}_{\mu\nu,\rho\sigma}(q)=-D^{(11)}_{\mu\nu,\rho\sigma}(q)$ (see Eq.~\eqref{temperaturepropagator}) has also been used, whereas $\mathbb{D}^{(11)}_{\mu\nu,\rho\sigma}(q;\beta)$ denotes the thermal contribution, for which
$\mathbb{D}^{(22)}_{\mu\nu,\rho\sigma}(q;\beta)=\mathbb{D}^{(11)}_{\mu\nu,\rho\sigma}(q;\beta)$.

The first term corresponds to the previously derived zero-temperature Bhabha scattering amplitude, modified only by a thermal factor involving the Fermi--Dirac distribution. This contribution is already well understood. The second and third terms, however, require a more careful analysis. Both $\mathbb{D}^{(11)}_{\mu\nu,\rho\sigma}(q;\beta)$ and $D^{(12)}_{\mu\nu,\rho\sigma}(q;\beta)$ contain the Bose-Einstein distribution multiplied by Dirac delta functions (see Eq.~\eqref{temperaturepropagator}). At first sight, one might expect these delta functions to eliminate the corresponding contributions once inserted into the effective potential. As shown below, however, this is not the case.

Within the effective potential, both contributions involve integrals of the form
\begin{align}
	I\propto\int \frac{d^3 \vec{q}}{(2\pi)^3}e^{i\vec{q}\cdot \vec{r}}n(\beta)\delta(q^2).
\end{align}
Using the properties of the Dirac delta function and writing the momentum-space measure as $\int d^3\vec{q}=\int_0^\pi\int 2\pi q^2\sin\theta\,dq\,d\theta$, we obtain
\begin{align}
	I=\frac{1}{(2\pi)^2}\int_0^\pi\int q^2\sin\theta\,e^{i\vec{q}\cdot\vec{r}\cos\theta}n(\beta)\left[\frac{\delta(\vec{q}-q_0)}{2q_0}+\frac{\delta(\vec{q}+q_0)}{2q_0}\right]dq\,d\theta,
\end{align}
which reduces to
\begin{align}
	I=\frac{1}{(2\pi)^2}\int_0^\pi q_0n(\beta)\sin\theta\cos\left(q_0\vec{r}\cos\theta\right)d\theta.
\end{align}
Evaluating the integral and recalling that
\begin{align}
	n(\beta)=\frac{1}{e^{\beta q_0}-1},
\end{align}
it is found that
\begin{align}
	I=\frac{\sin(q_0\vec{r})}{2\pi^2\vec{r}(e^{\beta q_0}-1)}.
\end{align}
In the non-relativistic limit, $q_0\rightarrow0$, so that $e^{\beta q_0}-1\approx\beta q_0$. Therefore,
\begin{align}
	I=\lim_{q_0\to0}\frac{\sin(q_0\vec{r})}{2\pi^2\beta q_0\vec{r}}\approx\frac{1}{2\pi^2\beta}.
\end{align}

Hence, the thermal contribution associated with the terms proportional to $\delta(q^2)$ is explicitly derived and shown to approach a finite constant. Substituting Eq.~\eqref{Mfinal} into the definition of the effective potential and using the previous result, the following expression is obtained
\begin{align}
	V(\vec{r};\beta)=-\frac{Gm^2}{\vec{r}}\tanh\left(\frac{\beta m}{2}\right)+\frac{8iGm^2}{\beta}\tanh^2\left(\frac{\beta m}{2}\right),
\end{align}
where the hyperbolic identities relating $U_\beta$ and $V_\beta$ have been used.

This result represents the finite-temperature gravitational potential associated with bringing a particle from infinity to a given point in space, namely the long-range gravitational interaction in the presence of a thermal bath. At first sight, the appearance of an imaginary contribution may seem unusual. However, the time evolution of a quantum state is given by $\varphi(t)\propto e^{-iEt}$. Substituting the effective potential into this expression shows that the imaginary part generates a real exponential damping factor, describing the dissipation of the quantum state. Thus, while the real part of the effective potential determines the gravitational interaction between particles immersed in the thermal bath, the imaginary contribution quantifies the rate at which the thermal environment suppresses the coherence of the interacting quantum state.

\vspace{0.2cm}
\begin{figure}[htpb]
	\centering
	\includegraphics[width=0.98\textwidth]{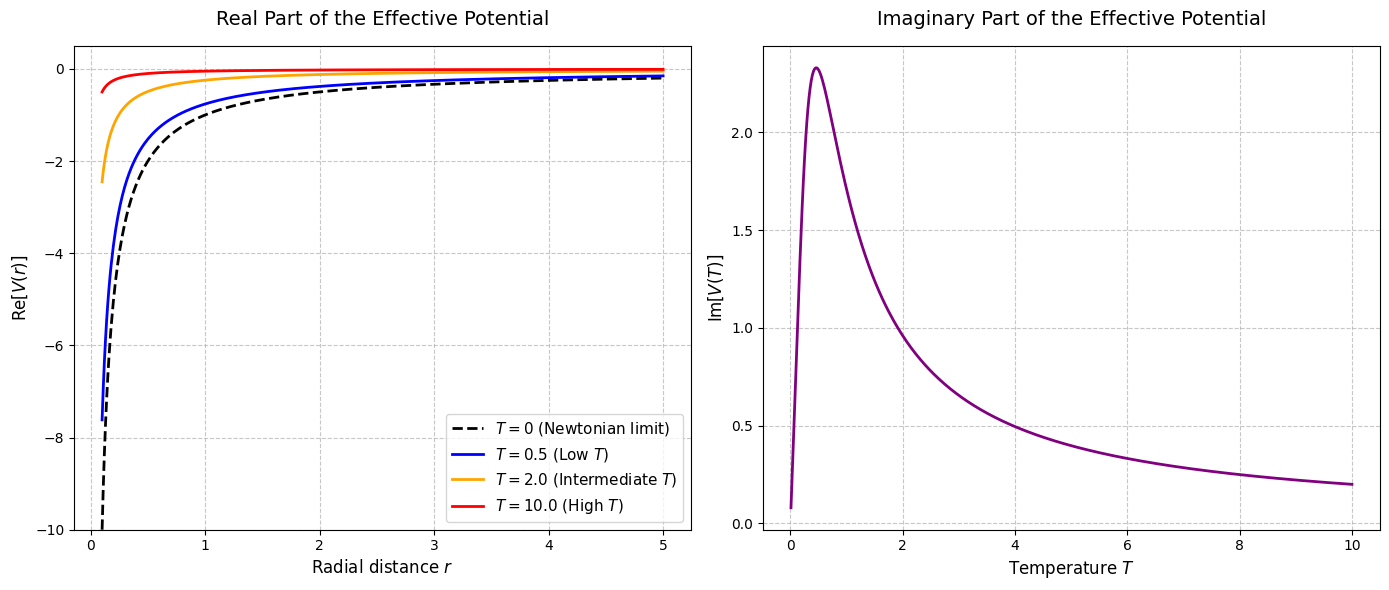}
	\caption{Behavior of the finite-temperature effective potential in the Weyl GEM framework. The left panel shows the real part of the potential (spatial interaction) as a function of the radial distance $r$ for different temperature regimes. The right panel displays the imaginary contribution (thermal dissipation rate) as a function of the temperature $T$.}
	\label{fig:effective_potential_thermal}
\end{figure}

The behavior of both components is illustrated in Fig.~\ref{fig:effective_potential_thermal}. For illustrative purposes, the graphical representation is plotted in natural units ($G=1$, $m=1$), aiming to highlight the phenomenological behavior of the potential rather than its absolute magnitude. As shown in the left panel, in the zero-temperature limit ($\beta\rightarrow\infty$), we have $\tanh(\beta m/2)\rightarrow1$, whereas the imaginary contribution vanishes, recovering the purely attractive Newtonian potential. Physically, this corresponds to the absence of a thermal bath, in which case the doubled Hilbert space effectively reduces to the ordinary quantum vacuum. Conversely, in the high-temperature limit ($\beta\rightarrow0$), we obtain $\tanh(\beta m/2)\rightarrow0$, causing the real part of the potential to vanish. As depicted by the progressive flattening of the curves, in a sufficiently hot plasma the gravitational interaction mediated by the GEM field is completely screened, rendering the medium effectively opaque to the exchange of virtual gravitons.

The right panel of Fig.~\ref{fig:effective_potential_thermal} displays the imaginary contribution. Since this term is spatially uniform, it is plotted only as a function of $T$ to represent the thermal dissipation rate. The resulting non-monotonic behavior reflects the competition between two distinct mechanisms. At intermediate temperatures, the increasing population of thermal excitations enhances the dissipation. At sufficiently high temperatures, however, phase-space saturation governed by the Fermi-Dirac distribution suppresses the interaction between external particles. As the momentum states become progressively occupied by the particles of the thermal bath, fewer accessible states remain available for the scattering process, reducing the effectiveness of virtual graviton exchange. Consequently, the dissipation rate also decreases in the extreme high-temperature regime.

A qualitatively similar suppression of interactions is also encountered in finite-temperature QED, although there it is usually associated with Debye screening generated through loop corrections. The mechanism discussed here is fundamentally different. The screening already appears at tree level, despite the mediating field remaining massless, indicating that it is not associated with an effective thermal mass. Instead, it appears to originate from the combined effects of the thermalized vacuum, the specific scattering process considered, and the non-relativistic approximation adopted in the calculation. Furthermore, previous studies of scattering processes within the GEM framework \cite{evangelista2025gravitational}, in which the complete set of Feynman diagrams is taken into account, have shown that thermal effects may instead enhance the interaction strength. This suggests that thermal corrections in GEM are process dependent rather than universal, with their qualitative behavior determined by the particular scattering channel under consideration.

\section{Conclusions}\label{V}

In this work, the Weyl GEM framework was employed to calculate the effective potential associated with Bhabha scattering. Weyl GEM is one of the three formulations of Gravitoelectromagnetism (GEM) and is based on the decomposition of the Weyl curvature tensor into its gravitoelectric and gravitomagnetic components. To preserve the structural analogy with Maxwell's electromagnetism, the theory adopts a scalar-generated gauge symmetry, which introduces additional propagation modes. Fortunately, these extra modes do not contribute to physical scattering amplitudes. As a result, the effective potential between interacting fermions mediated by the $A_{\mu\nu}$ field exactly reproduces the Newtonian gravitational potential, demonstrating that the Weyl GEM formalism provides an efficient alternative framework for describing weak-field gravitational interactions while remaining consistent with the predictions of General Relativity.

The analysis was then extended to finite temperature through the Thermo Field Dynamics (TFD) formalism. TFD is a real-time approach that incorporates thermal effects by doubling the Hilbert space through the introduction of an identical auxiliary, or tilde, space. Within this framework, thermal corrections to the effective potential naturally arise through the Fermi-Dirac and Bose-Einstein statistical distributions. In the low-temperature limit, $T\rightarrow0$, the finite-temperature potential continuously reduces to the Newtonian potential obtained at zero temperature. In contrast, in the high-temperature limit, the real part of the effective potential vanishes, indicating a thermal screening of the gravitational interaction. This behavior is interpreted as a consequence of the thermalized vacuum, where the increasing occupation of momentum states by thermal excitations progressively suppresses the exchange of virtual gravitons between external particles. The imaginary contribution to the effective potential was also analyzed and interpreted as a dissipative effect associated with the thermal environment, introducing damping into the quantum evolution of the interacting system.

Finally, it is important to emphasize that these conclusions are specific to the scattering process considered in this work. The thermal screening discussed here emerges already at tree level, distinguishing it from the Debye screening mechanism commonly encountered in finite-temperature QED. Since the effective potential was derived from a particular scattering channel, different interaction processes or the inclusion of additional Feynman diagrams may lead to quantitatively or even qualitatively different thermal corrections. This suggests that thermal effects within the Weyl GEM framework are intrinsically process dependent and deserve further investigation in more general scattering scenarios.

\section*{Acknowledgments}

This work by A. F. S. is partially supported by National Council for Scientific and Technological
Development - CNPq project No. 312406/2023-1. L. A. S. E. thanks CAPES for financial support.

\section*{Data Availability Statement}

No Data associated in the manuscript.

\section*{Conflicts of Interest}

No conflict of interests in this paper.


\global\long\def\link#1#2{\href{http://eudml.org/#1}{#2}}
 \global\long\def\doi#1#2{\href{http://dx.doi.org/#1}{#2}}
 \global\long\def\arXiv#1#2{\href{http://arxiv.org/abs/#1}{arXiv:#1 [#2]}}
 \global\long\def\arXivOld#1{\href{http://arxiv.org/abs/#1}{arXiv:#1}}


\end{document}